\documentclass[12pt,english,aps,reprint,showpacs,prl,amsmath,amssymb,floatfix,superscriptaddress]{revtex4-1}
\usepackage[T1]{fontenc}
\usepackage[latin9]{inputenc}
\usepackage{color}
\usepackage{babel}
\usepackage{amsmath}
\usepackage{amssymb}
\usepackage{graphicx}
\usepackage[unicode=true,
 bookmarks=true,bookmarksnumbered=false,bookmarksopen=false,
 breaklinks=false,pdfborder={0 0 1},backref=false,colorlinks=false]
 {hyperref}
\hypersetup{pdftitle={Damping of magnetization dynamics by phonon pumping},
 pdfauthor={Simon Streib}}

\makeatletter
\usepackage{placeins}

\makeatother

\begin{document}

\title{Damping of magnetization dynamics by phonon pumping}

\author{Simon Streib}

\affiliation{Kavli Institute of NanoScience, Delft University of Technology, Lorentzweg
1, 2628 CJ Delft, The Netherlands}

\author{Hedyeh Keshtgar}

\affiliation{Institute for Advanced Studies in Basic Science, 45195 Zanjan, Iran}

\author{Gerrit E. W. Bauer}

\affiliation{Kavli Institute of NanoScience, Delft University of Technology, Lorentzweg
1, 2628 CJ Delft, The Netherlands}

\affiliation{Institute for Materials Research \& WPI-AIMR \& CSRN, Tohoku University,
Sendai 980-8577, Japan}

\date{July 11, 2018}
\begin{abstract}
We theoretically investigate pumping of phonons by the dynamics of
a magnetic film into a non-magnetic contact. The enhanced damping
due to the loss of energy and angular momentum shows interference
patterns as a function of resonance frequency and magnetic film thickness
that cannot be described by viscous (``Gilbert'') damping. The phonon
pumping depends on magnetization direction as well as geometrical
and material parameters and is observable, e.g., in thin films of
yttrium iron garnet on a thick dielectric substrate. 
\end{abstract}
\maketitle
The dynamics of ferromagnetic heterostructures is at the root of devices
for information and communication technologies \cite{Bader2010,Kruglyak2010,Pulizzi2012,Chumak2015,Nikitov2015}.
When a normal metal contact is attached to a ferromagnet, the magnetization
dynamics drives a spin current through the interface. This effect
is known as spin pumping and can strongly enhance the (Gilbert) viscous
damping in ultra-thin magnetic films \cite{Tserkovnyak2002,Tserkovnyak2005,Kapelrud2013}.
Spin pumping and its (Onsager) reciprocal, the spin transfer torque
\cite{Berger1996,Slonczewski1996}, are crucial in spintronics, as
they allow electric control and detection of magnetization dynamics.
When a magnet is connected to a non-magnetic insulator instead of
a metal, angular momentum cannot leave the magnet in the form of electronic
or magnonic spin currents, but they can do so in the form of phonons.
Half a century ago it was reported \cite{Boemmel1959,Pomerantz1961}
and explained \cite{Comstock1963,Seavey1965,Kobayashi1971,Kobayashi1972}
that magnetization dynamics can generate phonons by magnetostriction.
More recently, the inverse effect of magnetization dynamics excited
by surface acoustic waves (SAWs) has been studied \cite{Weiler2011,Dreher2012,Gowtham2015,Li2017}
and found to generate spin currents in proximity normal metals \cite{Uchida2011,Weiler2012}.
The emission and detection of SAWs was combined in one and the same
device \cite{Bhuktare2017a,Bhuktare2017b}, and adiabatic transformation
between magnons and phonons was observed in inhomogeneous magnetic
fields \cite{Holanda2018}. The angular momentum of phonons \cite{Levine1962,McLellan1988}
has recently come into focus again in the context of the Einstein-de
Haas effect \cite{Zhang2014} and spin-phonon interactions in general
\cite{Garanin2015}. The interpretation of the phonon angular momentum
in terms of orbital and spin contributions \cite{Garanin2015} has
been challenged \cite{Tiwari2017}, a discussion that bears similarities
with the interpretation of the photon angular momentum \cite{Leader2016}.
In our opinion this distinction is rather semantic since not required
to arrive at concrete results. A recent quantum theory of the dynamics
of a magnetic impurity \cite{Mentink2018} predicts a broadening of
the electron spin resonance and a renormalized g-factor by coupling
to an elastic continuum via the spin-orbit interaction, which appears
to be related to the enhanced damping and effective gyromagnetic ratio
discussed here.

A phonon current generated by magnetization dynamics generates damping
by carrying away angular momentum and energy from the ferromagnet.
While the phonon contribution to the bulk Gilbert damping has been
studied theoretically \cite{Kobayashi1971b,Kobayashi1973b,Kobayashi1973,Rossi2005,Widom2010,Vittoria2010},
the damping enhancement by interfaces to non-magnetic substrates or
overlayers has to our knowledge not been addressed before. Here we
present a theory of the coupled lattice and magnetization dynamics
of a ferromagnetic film attached to a half-infinite non-magnet, which
serves as an ideal phonon sink. We predict, for instance, significantly
enhanced damping when an yttrium iron garnet (YIG) film is grown on
a thick gadolinium gallium garnet (GGG) substrate.
\begin{figure}
\begin{centering}
\includegraphics[scale=1.3]{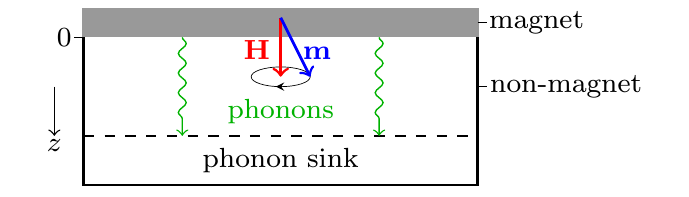}
\par\end{centering}
\caption{Magnetic film (shaded) with magnetization $\mathbf{m}$ attached to
a semi-infinite elastic material, which serves as an ideal phonon
sink. \label{fig:film}}
\end{figure}

We consider an easy-axis magnetic film with static external magnetic
field and equilibrium magnetization either normal (see Fig.~\ref{fig:film})
or parallel to the plane. The magnet is connected to a semi-infinite
elastic material. Magnetization and lattice are coupled by the magnetocrystalline
anisotropy and the magnetoelastic interaction, giving rise to coupled
field equations of motion in the magnet \cite{Abrahams1952,Kittel1953,Kittel1958,Kaganov1959}.
By matching these with the lattice dynamics in the non-magnet by proper
boundary conditions, we predict the dynamics of the heterostructure
as a function of geometrical and constitutive parameters. We find
that magnetization dynamics induced, e.g., by ferromagnetic resonance
(FMR) excites the lattice in the attached non-magnet. In analogy with
the electronic case we call this effect ``phonon pumping'' that
affects the magnetization dynamics. We consider only equilibrium magnetizations
that are normal or parallel to the interface, in which the pumped
phonons are pure shear waves that carry angular momentum. We note
that for general magnetization directions both shear and pressure
waves are emitted, however.

We consider a magnetic film (metallic or insulating) that extends
from $z=-d$ to $z=0$. It is subject to sufficiently high magnetic
fields $H_{0}$ such that magnetization is uniform, i.e. $\mathbf{M}(\mathbf{r})=\mathbf{M.}$
For in-plane magnetizations, $H_{0}>M_{s}$, where the magnetization
$M_{s}$ governs the demagnetizing field~\cite{Knuepfer2011}. The
energy of the magnet|non-magnet bilayer can be written

\begin{equation}
E=E_{T}+E_{el}+E_{Z}+E_{D}+E_{K}^{0}+E_{me},\label{eq:energy}
\end{equation}
which are integrals over the energy densities $\varepsilon_{X}(\mathbf{r})$.
The different contributions are explained in the following.

The kinetic energy density of the elastic motion reads
\begin{equation}
\varepsilon_{T}(\mathbf{r})=\begin{cases}
\frac{1}{2}\rho\dot{\mathbf{u}}^{2}(\mathbf{r}), & z>0\\
\frac{1}{2}\tilde{\rho}\dot{\mathbf{u}}^{2}(\mathbf{r}), & -d<z<0
\end{cases},
\end{equation}
and the elastic energy density \cite{Landau1970} 
\begin{equation}
\varepsilon_{el}=\begin{cases}
\frac{1}{2}\lambda\left(\sum_{\alpha}X_{\alpha\alpha}(\mathbf{r})\right)^{2}+\mu\sum_{\alpha\beta}X_{\alpha\beta}^{2}(\mathbf{r}), & z>0\\
\frac{1}{2}\tilde{\lambda}\left(\sum_{\alpha}X_{\alpha\alpha}(\mathbf{r})\right)^{2}+\tilde{\mu}\sum_{\alpha\beta}X_{\alpha\beta}^{2}(\mathbf{r}), & \!\!\!\!-d<z<0
\end{cases},\label{eq:elastic energy density}
\end{equation}
where $\alpha,\beta\in\{x,y,z\}$, $\lambda$ and $\mu$ are the Lamé
parameters and $\rho$ the mass density of the non-magnet. The tilded
parameters are those of the magnet. The strain tensor $X_{\alpha\beta}$
is defined in terms of the displacement fields $u_{\alpha}(\mathbf{r})$,
\begin{equation}
X_{\alpha\beta}(\mathbf{r})=\frac{1}{2}\left(\frac{\partial u_{\alpha}(\mathbf{r})}{\partial r_{\beta}}+\frac{\partial u_{\beta}(\mathbf{r})}{\partial r_{\alpha}}\right).
\end{equation}
$E_{Z}=-\mu_{0}V\mathbf{M}\cdot\mathbf{H}_{\mathrm{ext}}$ is the
Zeeman energy for $\mathbf{H}_{\mathrm{ext}}=\mathbf{H}_{\mathrm{0}}+\mathbf{h}(t)$,
where \textbf{$\mathbf{h}\left(t\right)$ }is time-dependent. $E_{D}=\frac{1}{2}\mu_{0}V\mathbf{M^{\mathit{T}}}\mathcal{D}\mathbf{M}$
is the magnetostatic energy with shape-dependent demagnetization tensor
$\mathcal{D}$ and $V$ the volume of the magnet. For a thin film
with $z$ axis along the surface normal $\mathbf{n}_{0}$, $\mathcal{D}_{zz}=1$
while the other components vanish. $E_{K}^{0}=K_{1}V\left(\mathbf{m}\times\mathbf{n}_{0}\right)^{2}$
is the uniaxial magnetocrystalline anisotropy in the absence of lattice
deformations, where $\mathbf{m}=\mathbf{M}/M_{s}$ and $K_{1}$ is
the anisotropy constant. The magnetoelastic energy $E_{me}$ couples
the magnetization to the lattice, as discussed in the following.

\textcolor{black}{The mag}netoelastic energy density can be expanded
as
\begin{align}
\varepsilon_{me}(\mathbf{r})=\frac{1}{M_{s}^{2}}\sum_{\alpha,\beta} & M_{\alpha}(\mathbf{r})M_{\beta}(\mathbf{r})\nonumber \\
 & \times\left[B_{\alpha\beta}X_{\alpha\beta}(\mathbf{r})+C_{\alpha\beta}\Omega_{\alpha\beta}(\mathbf{r})\right].\label{eq:magnetoelastic energy}
\end{align}
For an isotropic medium the magnetoelastic constants $B_{\alpha\beta}$
read \cite{Rueckriegel2014} 
\begin{align}
B_{\alpha\beta} & =\delta_{\alpha\beta}B_{\parallel}+(1-\delta_{\alpha\beta})B_{\perp}.
\end{align}
Rotational deformations as expressed by the tensor 
\begin{equation}
\Omega_{\alpha\beta}(\mathbf{r})=\frac{1}{2}\left(\frac{\partial u_{\alpha}(\mathbf{r})}{\partial r_{\beta}}-\frac{\partial u_{\beta}(\mathbf{r})}{\partial r_{\alpha}}\right)
\end{equation}
are often disregarded \cite{Abrahams1952,Kittel1953,Kittel1958,Kaganov1959,Kittel1949},
but lead to a position dependence of the easy axis $\mathbf{\mathbf{n}(\mathbf{r})}$
from the equilibrium value $\mathbf{n}_{0}=\mathbf{e}_{z}$ and an
anisotropy energy density \cite{Garanin1997,Jaafar2009,Garanin2015}
\begin{equation}
\varepsilon_{K}(\mathbf{r})=\frac{K_{1}}{M_{s}^{2}}\left[\mathbf{M}\times\mathbf{n}(\mathbf{r})\right]^{2}.
\end{equation}
To first order in the small deformation
\begin{equation}
\delta\mathbf{n}(\mathbf{r})=\mathbf{n}(\mathbf{r})-\mathbf{n}_{0}=\begin{pmatrix}\Omega_{xz}(\mathbf{r})\\
\Omega_{yz}(\mathbf{r})\\
0
\end{pmatrix},
\end{equation}
\begin{equation}
\varepsilon_{K}(\mathbf{r})=\varepsilon_{K}^{0}+2K_{1}\left(\mathbf{n}_{0}-m_{z}\mathbf{m}\right)\cdot\delta\mathbf{n}(\mathbf{r}).
\end{equation}
From $\Omega_{\alpha\beta}=-\Omega_{\beta\alpha}$ it follows that
(for non-chiral crystal structures) $C_{\alpha\beta}=-C_{\beta\alpha}$.
For the uniaxial anisotropy considered here $C_{xz}=C_{yz}=-K_{1}$.
The magnetoelastic coupling due to the magnetocrystalline anisotropy
thus contributes \cite{Garanin1997}
\begin{equation}
\varepsilon_{me}^{K}(\mathbf{r})=-\frac{2K_{1}}{M_{s}^{2}}M_{z}(\mathbf{r})\left[M_{x}(\mathbf{r})\Omega_{xz}(\mathbf{r})+M_{y}(\mathbf{r})\Omega_{yz}(\mathbf{r})\right].\label{eq:E_me^K}
\end{equation}
Pure YIG is magnetically very soft, so the magnetoelastic constants
are much larger than the anisotropy constant \cite{Gurevich1996,Hansen1974}
\begin{align}
B_{\parallel} & =3.48\times10^{5}\;\mathrm{J}\mathrm{m}^{-3},\quad B_{\perp}=6.96\times10^{5}\;\mathrm{J}\mathrm{m}^{-3},\nonumber \\
K_{1} & =-6.10\times10^{2}\;\mathrm{J\,m}^{-3},
\end{align}
but this ratio can be very different for other magnets. We find below
that for the Kittel mode dynamics both coupling processes cannot be
distinguished, even though they can characteristically affect the
magnon-phonon coupling for finite wave numbers.

The magnetization dynamics within the magnetic film is described by
the Landau-Lifshitz-Gilbert (LLG) equation \cite{Landau1935,Gilbert2004}
\begin{eqnarray}
\dot{\mathbf{m}} & = & -\gamma\mu_{0}\mathbf{m}\times\mathbf{H}_{\mathrm{eff}}+\boldsymbol{\mathbf{\tau}}_{m}^{(\alpha)},\label{eq:LLG equation}
\end{eqnarray}
where $-\gamma$ is the gyromagnetic ratio, the effective magnetic
field which includes the magnetoelastic coupling
\begin{equation}
\mathbf{H}_{\mathrm{eff}}=-\mathbf{\nabla}_{\mathbf{m}}E/(\mu_{0}VM_{s}),
\end{equation}
and the Gilbert damping torque \cite{Gilbert2004}
\begin{equation}
\boldsymbol{\tau}_{m}^{(\alpha)}=\alpha\mathbf{m}\times\dot{\mathbf{m}}.\label{eq:Gilbert damping}
\end{equation}
The equation of motion of the elastic continuum reads \cite{Landau1970}
\begin{equation}
\ddot{{\bf u}}({\bf r},t)=c_{t}^{2}\triangle{\bf u}({\bf r},t)+(c_{l}^{2}-c_{t}^{2}){\bf \nabla}\left[{\bf \nabla}\cdot{\bf u}({\bf r},t)\right],\label{eq:elastic equation of motion}
\end{equation}
with longitudinal and transverse sound velocities
\begin{equation}
c_{l}=\sqrt{\frac{\lambda+2\mu}{\rho}},\quad c_{t}=\sqrt{\frac{\mu}{\rho}},
\end{equation}
where elastic constants and mass density of non-magnet and magnet
can differ.

A uniform precession of the magnetization interacts with the lattice
deformation at the surfaces of the magnetic film \cite{Comstock1963,Seavey1965}
and at defects in the bulk. The present theory then holds when the
thickness of the magnetic film $d\ll\sqrt{A}$, where $A$ is the
cross section area. The Kittel mode induces lattice distortions that
are uniform in the film plane $u_{\alpha}(\mathbf{r})=u_{\alpha}(z)$
\cite{Seavey1965}. The elastic energy density is then affected by
shear waves only:
\begin{equation}
\varepsilon_{el}(z)=\begin{cases}
\frac{\mu}{2}\left(u_{x}'^{2}(z)+u_{y}'^{2}(z)\right), & z>0\\
\frac{\tilde{\mu}}{2}\left(u_{x}'^{2}(z)+u_{y}'^{2}(z)\right), & -d<z<0
\end{cases},
\end{equation}
where $u_{\alpha}'(z)=\partial u_{\alpha}(z)/\partial z$. The magnetic
field $\mathbf{H}_{\mathrm{ext}}=\begin{pmatrix}h_{x}(t), & h_{y}(t), & H_{0}\end{pmatrix}^{\mathrm{T}}$
with monochromatic drive $h_{x,y}(t)=\mathrm{Re}\left(h_{x,y}e^{-i\omega t}\right)$
and static component $H_{0}$ along the $z$ axis. At the FMR frequency
$\omega_{\perp}=\omega_{H}+\omega_{A}$ with $\omega_{H}=\gamma\mu_{0}H_{0}$
and $\omega_{A}=\gamma\left(2K_{1}/M_{s}-\gamma\mu M_{s}\right)$.
The equilibrium magnetization is perpendicular for $\omega_{\perp}>0$.
The magnetoelastic energy derived above then simplifies to
\begin{equation}
E_{me}^{z}=\frac{\left(B_{\perp}-K_{1}\right)A}{M_{s}}\sum_{\alpha=x,y}M_{\alpha}\left[u_{\alpha}(0)-u_{\alpha}(-d)\right],\label{eq: E^z}
\end{equation}
which results in surface shear forces $F_{\pm}(0)=-F_{\pm}(-d)=-\left(B_{\perp}-K_{1}\right)Am_{\pm}$,
with $F_{\pm}=F_{x}\pm iF_{y}$. These forces generate a stress or
transverse momentum current in the $z$ direction (see Supplemental
Material) 
\begin{equation}
j_{\pm}(z)=-\mu(z)u_{\pm}'(z),
\end{equation}
with $\mu(z)=\mu$ for $z>0$ and $\mu(z)=\tilde{\mu}$ for $-d<z<0$,
and $u_{\text{\ensuremath{\pm}}}=u_{x}\pm iu_{y}$, which is related
to the transverse momentum $p_{\pm}(z)=\rho\left(\dot{u}_{x}(z)\pm i\dot{u}_{y}(z)\right)$
by Newton's equation:
\begin{equation}
\dot{p}_{\pm}(z)=-\frac{\partial}{\partial z}j_{\pm}(z).
\end{equation}
The boundary conditions require momentum conservation and elastic
continuity at the interfaces, 
\begin{align}
j_{\pm}(-d) & =\left(B_{\perp}-K_{1}\right)m_{\pm},\\
j_{\pm}(0^{+})-j_{\pm}(0^{-}) & =-\left(B_{\perp}-K_{1}\right)m_{\pm},\\
u_{\pm}(0^{+}) & =u_{\pm}(0^{-}).
\end{align}
We treat the magnetoelastic coupling as a small perturbation and therefore
we approximate the magnetization $m_{\pm}$ entering the above boundary
conditions as independent of the lattice displacement $u_{\pm}$.
The loss of angular momentum (see Supplemental Material) affects the
magnetization dynamics in the LLG equation in the form of a torque,
which we derive from the magnetoelastic energy (\ref{eq: E^z}),
\begin{align}
\left.\dot{m}_{\pm}\right|_{me} & =\pm i\frac{\omega_{c}}{d}\left[u_{\pm}(0)-u_{\pm}(-d)\right]\nonumber \\
 & =\pm i\omega_{c}\mathrm{Re}(v)m_{\pm}\mp\omega_{c}\mathrm{Im}(v)m_{\pm},
\end{align}
where $\omega_{c}=\gamma\left(B_{\perp}-K_{1}\right)/M_{s}$ (for
YIG: $\omega_{c}=8.76\times10^{11}\;\mathrm{s}^{-1}$) and $v=\left[u_{\pm}(0)-u_{\pm}(-d)\right]/(dm_{\pm})$.
We can distinguish an effective field
\begin{equation}
\mathbf{H}_{me}=\frac{\omega_{c}}{\gamma\mu_{0}}\mathrm{Re}(v)\mathbf{e}_{z},\label{eq:H_ms}
\end{equation}
and a damping coefficient 
\begin{equation}
\alpha_{me}^{(\perp)}=-\frac{\omega_{c}}{\omega}\mathrm{Im}\,v.\label{eq:alpha_me}
\end{equation}
The latter can be compared with the Gilbert damping constant $\alpha$
that enters the linearized equation of motion as
\begin{equation}
\left.\dot{m}_{\pm}\right|_{\alpha}=\pm i\alpha\dot{m}_{\pm}=\pm\alpha\omega m_{\pm}.
\end{equation}
With the ansatz
\begin{equation}
u_{\pm}(z,t)=\begin{cases}
C_{\pm}e^{ikz-i\omega t}, & z>0\\
D_{\pm}e^{i\tilde{k}z-i\omega t}+E_{\pm}e^{-i\tilde{k}z-i\omega t}, & -d<z<0
\end{cases},
\end{equation}
we obtain
\begin{equation}
v=\frac{M_{s}\omega_{c}}{\omega\gamma d\tilde{\rho}\tilde{c}_{t}}\frac{2\left[\cos(\tilde{k}d)-1\right]-i\frac{\rho c_{t}}{\tilde{\rho}\tilde{c}_{t}}\sin(\tilde{k}d)}{\sin(\tilde{k}d)+i\frac{\rho c_{t}}{\tilde{\rho}\tilde{c}_{t}}\cos(\tilde{k}d)},
\end{equation}
and the damping coefficient for perpendicular magnetization
\begin{equation}
\alpha_{me}^{(\perp)}=\mathrm{\left(\frac{\omega_{c}}{\omega}\right)^{2}}\frac{M_{s}}{\gamma d\tilde{\rho}\tilde{c}_{t}}\frac{\rho c_{t}}{\tilde{\rho}\tilde{c}_{t}}\frac{4\sin^{4}\left(\frac{\tilde{k}d}{2}\right)}{\sin^{2}(\tilde{k}d)+\left(\frac{\rho c_{t}}{\tilde{\rho}\tilde{c}_{t}}\right)^{2}\cos^{2}(\tilde{k}d)},\label{eq:damping result}
\end{equation}
where $\omega=c_{t}k=\tilde{c}_{t}\tilde{k}$. The oscillatory behavior
of the damping $\alpha_{me}^{(\perp)}$ comes from the interference
of the elastic waves that are generated at the top and bottom surfaces
of the magnetic film. When they constructively (destructively) interfere
at the FMR frequency, the damping is enhanced (suppressed), because
the magnon-phonon coupling and phonon emission are large (small).

When $\rho c_{t}\ll\tilde{\rho}\tilde{c}_{t}$ (soft substrate) or
when acoustic impedances are matched ($\rho c_{t}=\tilde{\rho}\tilde{c}_{t}$),
damping at the resonance $\tilde{k}d=(2n+1)\pi$ with $n\in\mathbb{N}_{0}$
\cite{Seavey1965} simplifies to
\begin{equation}
\alpha_{me}^{(\perp)}\rightarrow\mathrm{\left(\frac{\omega_{c}}{\omega}\right)^{2}}\frac{4M_{s}}{\gamma d\rho c_{t}}.
\end{equation}
When $\rho c_{t}\gg\tilde{\rho}\tilde{c}_{t}$ (hard substrate), the
magnet is acoustically pinned at the interface and the acoustic resonances
are at $\tilde{k}d=(2n+1)\pi/2$ \cite{Seavey1965} with
\begin{equation}
\alpha_{me}^{(\perp)}\rightarrow\mathrm{\left(\frac{\omega_{c}}{\omega}\right)^{2}}\frac{M_{s}}{\gamma d\tilde{\rho}\tilde{c}_{t}}\frac{\rho c_{t}}{\tilde{\rho}\tilde{c}_{t}}.
\end{equation}

In contrast to Gilbert damping, $\alpha_{me}^{(\perp)}$ depends on
the frequency and vanishes in the limits $\omega\to0$ and $\omega\to\infty$.
Therefore, it does not obey the LLG phenomenology and in the non-linear
regime does not simply enhance $\alpha$ in Eq.~(\ref{eq:Gilbert damping}).
The magnetization damping $\alpha_{0}$ in bulk magnetic insulators,
on the other hand, is usually of the Gilbert type. It is caused by
phonons as well, but not necessarily the magnetoelastic coupling.
A theory of Gilbert damping \cite{Vittoria2010} assumes a bottleneck
process by sound wave attenuation, which appears realistic for magnets
with high acoustic quality such as YIG. In the present phonon pumping
model, energy and angular momentum is lost by the emission of sound
waves into an attached perfect phonon wave guide, so the pumping process
dominates. Such a scenario could also dominate the damping in magnets
in which the magnetic quality is relatively higher than the acoustic
one. 

When the field is rotated to $\mathbf{H}_{\mathrm{ext}}=\begin{pmatrix}h_{x}(t), & H_{0}, & h_{z}(t)\end{pmatrix}^{\mathrm{T}}$,
the equilibrium magnetization is in the in-plane $y$ direction and
the magnetoelastic energy couples only to the strain $u_{y}$,
\begin{equation}
E_{me}^{y}=\frac{\left(B_{\perp}-K_{1}\right)A}{M_{s}}M_{z}\left[u_{y}(0)-u_{y}(-d)\right].
\end{equation}
The FMR frequency for in-plane magnetization $\omega_{\parallel}=\omega_{H}\sqrt{1-\omega_{A}/\omega_{H}}$
with $\omega_{A}<\omega_{H}$. The magnetoelastic coupling generates
again only transverse sound waves. The linearized LLG equation including
the phononic torques reads now 
\begin{align}
\dot{m}_{x} & =(\omega_{H}+\omega_{me})m_{z}-\gamma\mu_{0}h_{z}-\omega_{A}m_{z}\nonumber \\
 & \phantom{=}+(\alpha+\alpha_{me})\dot{m}_{z},\\
\dot{m}_{z} & =-\omega_{H}m_{x}+\gamma\mu_{0}h_{x}-\alpha\dot{m}_{x},
\end{align}
where $\alpha_{me}$ is given by Eq.~(\ref{eq:alpha_me}) and $\omega_{me}=\gamma\mu_{0}H_{me}$
with effective field $H_{me}=\mathbf{H}_{me}\cdot\mathbf{e}_{z}$
given by Eq.~(\ref{eq:H_ms}). Both $H_{me}$ and $\alpha_{me}$
contribute only to $\dot{m}_{x}$. The phonon pumping is always less
efficient for the in-plane configuration:
\begin{equation}
\alpha_{me}^{(\parallel)}=\frac{1}{1+(\omega_{\parallel}/\omega_{H})^{2}}\alpha_{me}^{(\perp)}.
\end{equation}

As an example, we insert parameters for a thin YIG film on a semi-infinite
gadolinium gallium garnet (GGG) substrate at room temperature. We
have chosen YIG because of its low intrinsic damping and high quality
interface to the GGG substrate. Substantially larger magnetoelastic
coupling in other materials should be offset against generally larger
bulk damping. For GGG, $\rho=7.07\times10^{3}\;\mathrm{kg}\,\mathrm{m}^{-3}$,
$c_{l}=6411\;\mathrm{m\,s^{-1}}$, and $c_{t}=3568\;\mathrm{m\,s^{-1}}$
\cite{Kleszczewski1988}. For YIG, $M_{s}=1.4\times10^{5}\;\mathrm{A\,m^{-1}}$,
$\gamma=1.76\times10^{11}\;\mathrm{s}^{-1}\,\mathrm{T}^{-1}$, $\tilde{\rho}=5170\;\mathrm{kg}\,\mathrm{m}^{-3}$,
$\tilde{c}_{l}=7209\;\mathrm{m\,s^{-1}}$, $\tilde{c}_{t}=3843\;\mathrm{m\,s^{-1}}$,
and $\omega_{c}=8.76\times10^{11}\;\mathrm{s}^{-1}$ \cite{Hansen1974,Gurevich1996}.
The ratio of the acoustic impedances $\tilde{\rho}\tilde{c}_{t}/\rho c_{t}=0.79$.
\begin{figure}
\begin{centering}
\includegraphics{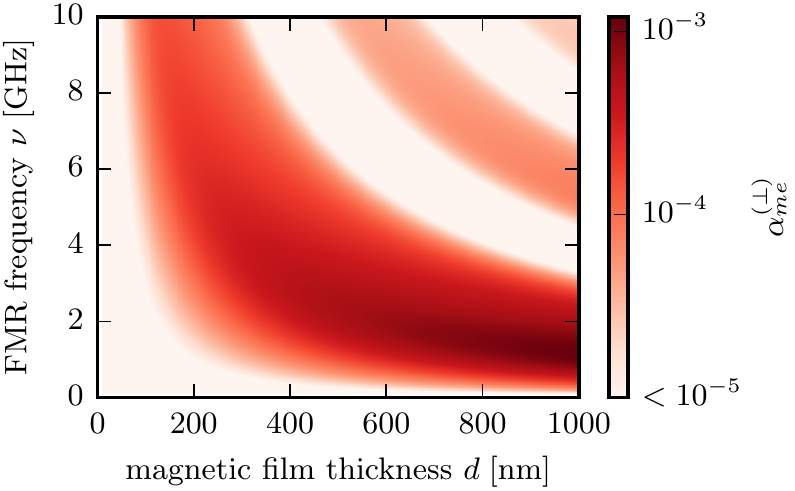}
\par\end{centering}
\caption{Damping enhancement $\alpha_{me}^{(\perp)}$ by phonon pumping in
a YIG film on a semi-infinite GGG substrate, as given by Eq.~(\ref{eq:damping result}).\label{fig:alpha ms}}
\end{figure}
The damping enhancement $\alpha_{me}^{(\perp)}$ is shown in Fig.~\ref{fig:alpha ms}
over a range of FMR frequencies and film thicknesses. The FMR frequencies
$\omega_{\perp}=\omega_{H}+\omega_{A}$ and $\omega_{\parallel}=\omega_{H}\sqrt{1-\omega_{A}/\omega_{H}}$
for the normal and in-plane configurations are tunable by the static
magnetic field component $H_{0}$ via $\omega_{H}=\gamma\mu_{0}H_{0}$.
The damping enhancement peaks at acoustic resonance frequencies $\nu\approx n\tilde{c}_{t}/(2d)$.
The counter-intuitive result that the damping increases for thicker
films can be understood by the competition between the magnetoelastic
effect that increases with thickness at the resonances and wins against
the increase in total magnetization. However, with increasing thickness
the resonance frequencies decrease below a minimum value at which
FMR can be excited. For a fixed FMR frequency $\alpha_{me}$$\rightarrow0$
for $d\rightarrow\infty.$ For comparison, the Gilbert damping in
nanometer thin YIG films is of the order $\alpha\sim10^{-4}$ \cite{Chang2014}
which is larger than corresponding values for single crystals. We
conclude that the enhanced damping is at least partly caused by interaction
with the substrate and not by a reduced crystal quality.

\textcolor{black}{The resonances in the figures are very broad because
the $\rho c_{t}\approx\tilde{\rho}\tilde{c}_{t}$ implies very strong
coupling of the discrete phonons in the thin magnetic layer with the
phonon continuum in the substrate. When an acoustic mismatch is introduced,
the broad peaks increasingly sharpen, reflecting the increased lifetime
of the magnon polaron resonances in the magnet. }

\begin{figure}
\begin{centering}
\includegraphics{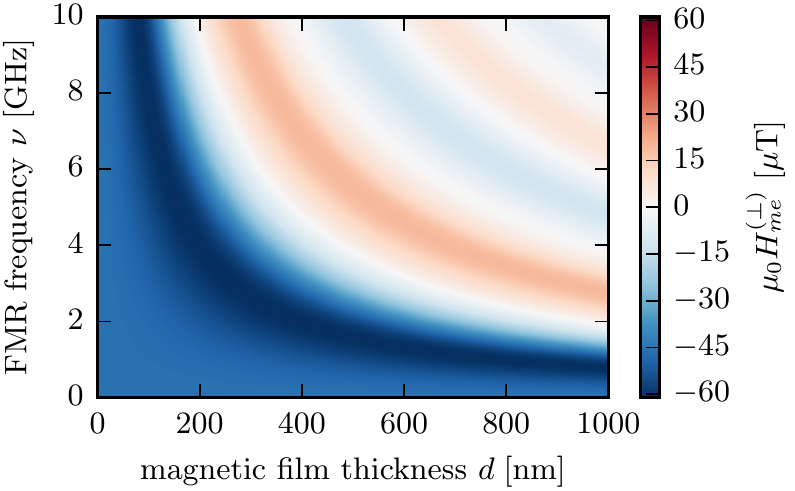}
\par\end{centering}
\caption{Effective field $H_{me}^{(\perp)}$ generated by the magnetoelastic
generation of phonons in a YIG film on a semi-infinite GGG substrate,
as given by Eq.~(\ref{eq:H_ms}).\label{fig:field}}
\end{figure}

The frequency dependent effective magnetic field $H_{me}^{(\perp)}$
is shown in Fig.~\ref{fig:field}. The frequency dependence of $H_{me}^{(\perp)}$
implies a weak frequency dependence of the effective gyromagnetic
ratio
\begin{equation}
\gamma_{\mathrm{eff}}^{(\perp)}=\gamma\left(1+\frac{\gamma\mu_{0}H_{me}^{(\perp)}}{\omega}\right).
\end{equation}
In the limit of vanishing film thickness, $\mu_{0}H_{me}^{(\perp)}\to-(B_{\perp}-K_{1})^{2}/(M_{s}\tilde{\mu})$.

We assumed that the non-magnet is an ideal phonon sink, which means
that injected sound waves do not return. In the opposite limit in
which the phonons cannot escape, i.e. when the substrate is a thin
film with high acoustic quality, the additional damping vanishes.
This can be interpreted in terms of a phonon accumulation that, when
allowed to thermalize, generates a phonon chemical potential and/or
non-equilibrium temperature. The non-equilibrium thermodynamics of
phonons in magnetic nanostructures is subject of our ongoing research.

The damping enhancement by phonons may be compared with that from
electronic spin pumping \cite{Tserkovnyak2002,Tserkovnyak2005,Kapelrud2013},

\begin{equation}
\alpha_{sp}=\frac{\gamma\hbar}{4\pi dM_{s}}\frac{h}{e^{2}}g,
\end{equation}
which is inversely proportional to the thickness $d$ of the magnetic
film and does not depend on the FMR frequency, i.e. obeys the LLG
phenomenology. Here, $g$ is the spin mixing conductance per unit
area at the interface. While phonons can be pumped into any elastic
material, spin pumping requires an electrically conducting contact.
With a typical value of $hg/e^{2}\sim10^{18}\;\mathrm{m}^{-2}$ the
damping enhancement of YIG on platinum is $\alpha_{sp}\sim10^{-2}\;\mathrm{nm}/d.$
The physics is quite different, however, since $\alpha_{sp},$ in
contrast to $\alpha_{me}$, does not require coherence over the interface.

In conclusion, the pumping of phonons by magnetic anisotropy and magnetostriction
causes a frequency-dependent contributions to the damping and effective
field of the magnetization dynamics. The generation of phonons by
magnetic precession can cause significant damping in a magnetic film
when grown on an insulating, non-magnetic substrate and partly explains
the increased damping invariably observed for thinner films. The implications
of further reaching ramifications, such as phonon-induced dynamic
exchange interactions, phonon accumulations and phonon spin Seebeck
effect require additional research. 
\begin{acknowledgments}
This work is financially supported by the Nederlandse Organisatie
voor Wetenschappelijk Onderzoek (NWO) as well as by the Grant-in-Aid
for Scientific Research on Innovative Area, \textquotedblright Nano
Spin Conversion Science\textquotedblright{} (Grant No. 26103006).
H. K. acknowledges support from the Iran Science Elites Federation.
We acknowledge useful discussions with Yaroslav Blanter, Rembert Duine,
Akashdeep Kamra, Eiji Saitoh, and Sanchar Sharma. 
\end{acknowledgments}

\bibliographystyle{apsrev4-1}

\onecolumngrid \clearpage \twocolumngrid \begin{center} \textbf{Supplemental Material} \end{center}
\setcounter{equation}{0} \setcounter{figure}{0} \setcounter{table}{0} \setcounter{page}{1} \makeatletter \renewcommand{\theequation}{S\arabic{equation}} \renewcommand{\thefigure}{S\arabic{figure}} 

In Secs.~\ref{sec:Angular-momentum} and \ref{sec:Transverse-momentum}
of this Supplement we derive the angular and transverse momentum of
transverse elastic waves and the corresponding momentum currents.
In Sec.~\ref{sec:Sandwiched-magnet} we give results for a magnet
sandwiched between two non-magnets and in Sec.~\ref{sec:Thin-beams}
we present a theory for the magnetization damping enhancement from
pumping flexural waves into a thin beam.

\section{Angular momentum\label{sec:Angular-momentum}}

The magnetization $\mathbf{M}=M_{s}\mathbf{m}$ of an uniformly magnetized
magnet with saturation magnetization $M_{s}$ and volume $V$ is associated
with the angular momentum
\begin{equation}
\mathbf{S}=-\frac{M_{s}V}{\gamma}\mathbf{m},
\end{equation}
where $-\gamma$ is the gyromagnetic ratio. The angular momentum density
relative to the origin of an elastic body with displacement field
$\mathbf{u}(\mathbf{r},t)$ and constant mass density $\rho$ reads
\begin{equation}
\mathbf{l}(t)=\rho\left(\mathbf{r}+\mathbf{u}(\mathbf{r},t)\right)\times\dot{\mathbf{u}}(\mathbf{r},t).
\end{equation}
With uniaxial anisotropy axis along $z$, FMR generates the transverse
elastic wave

\begin{equation}
\mathbf{u}(z,t)=\mathrm{Re}\left[\begin{pmatrix}u_{x}\\
u_{y}\\
0
\end{pmatrix}e^{ikz-i\omega t}\right],\label{eq:transverse elastic wave}
\end{equation}
with dispersion relation $\omega=c_{t}k$. Defining the time average
\begin{equation}
\left\langle f(t)\right\rangle =\lim_{T\rightarrow\infty}\frac{1}{T}\int_{0}^{T}dt\,f(t),
\end{equation}
 $\mathbf{\left\langle l\right\rangle =}\left(0,0,\left\langle l_{z}\right\rangle \right)$
can be expressed as 
\begin{align}
\left\langle l_{z}\right\rangle  & =\rho\left\langle u_{x}\dot{u}_{y}-\dot{u}_{x}u_{y}\right\rangle \nonumber \\
 & =-\frac{\rho\omega}{4}\left(\left|u_{+}\right|^{2}-\left|u_{-}\right|^{2}\right),
\end{align}
where $u_{\pm}=u_{x}\pm iu_{y}$ and where we used
\begin{equation}
\left\langle \mathrm{Re}(ae^{-i\omega t})\mathrm{Re}(be^{-i\omega t})\right\rangle =\frac{1}{2}\mathrm{Re}\left(a^{*}b\right).
\end{equation}
The non-magnet harbors a constant phonon angular momentum density
in the $z$-direction, which implies presence of a phonon angular
momentum current $A\left\langle j_{l}^{z}\right\rangle $ at the interface
to the magnet with area $A$ that is absorbed at the phonon sink.
In our model the angular momentum loss rate of the magnet by phonon
pumping $\langle\dot{S}_{z}|_{me}\rangle=-A\langle j_{l}^{z}\rangle$
and
\begin{equation}
\left\langle \left.\dot{S}_{z}\right|_{me}\right\rangle =\left\langle \left\{ S_{z},E_{me}^{z}\right\} \right\rangle =\frac{M_{s}V\alpha_{me}^{(\perp)}\omega}{4\gamma}\left(\left|m_{+}\right|^{2}-\left|m_{-}\right|^{2}\right)
\end{equation}
where $\{,\}$ is the Poisson bracket and $E_{me}^{z}$ the magnetoelastic
energy Eq.~(\ref{eq: E^z}) in the main text. From Eq.~(\ref{eq:damping result})
and
\begin{equation}
u_{\pm}=Cm_{\pm},
\end{equation}
with
\begin{equation}
C=\frac{\left(B_{\perp}-K_{1}\right)\left[\cos(\tilde{k}d)-1\right]}{ik\mu\cos(\tilde{k}d)+\tilde{k}\tilde{\mu}\sin(\tilde{k}d)},
\end{equation}
we obtain the relation
\begin{equation}
\left\langle j_{l}^{z}\right\rangle =c_{t}\left\langle l_{z}\right\rangle ,
\end{equation}
which agrees with the simple physical picture of an elastic wave carrying
away its angular momentum density $\left\langle l_{z}\right\rangle $
with transverse sound velocity $c_{t}$.

\section{Transverse momentum\label{sec:Transverse-momentum}}

For the transverse elastic wave (\ref{eq:transverse elastic wave})
in a magnet extending from $z=z_{0}$ to $z=z_{1}$ (with $z_{1}>z_{0}$),
the time derivative of the transverse momentum $P_{\pm}=P_{x}\pm iP_{y}$
reads 
\begin{align}
\dot{P}_{\pm} & =\rho\int_{V}d^{3}r\,\ddot{u}_{\pm}(z,t)\nonumber \\
 & =\mu A\left[u_{\pm}'(z_{1},t)-u_{\pm}'(z_{0},t)\right].
\end{align}
The change of momentum can be interpreted as a transverse momentum
current density $j_{\pm}(z_{0})=-\mu u_{\pm}'(z_{0})$ flowing into
the magnet at $z_{0}$ and a current $j_{\pm}(z_{1})=-\mu u_{\pm}'(z_{1})$
flowing out at $z_{1}$. The momentum current is related to the transverse
momentum density $p_{\pm}(z)=\rho\dot{u}_{\pm}(z)$ by
\begin{equation}
\dot{p}_{\pm}(z)=-\frac{\partial}{\partial z}j_{\pm}(z),
\end{equation}
which confirms that
\begin{equation}
j_{\pm}(z,t)=-\mu u_{\pm}'(z,t).
\end{equation}
The instantaneous conservation of transverse momentum is a boundary
conditions at the interface. Its time average $\left\langle j_{\pm}\right\rangle =0$,
but the associated angular momentum along $z$ is finite, as shown
above. 

\section{Sandwiched magnet\label{sec:Sandwiched-magnet}}

\begin{figure}
\begin{centering}
\includegraphics{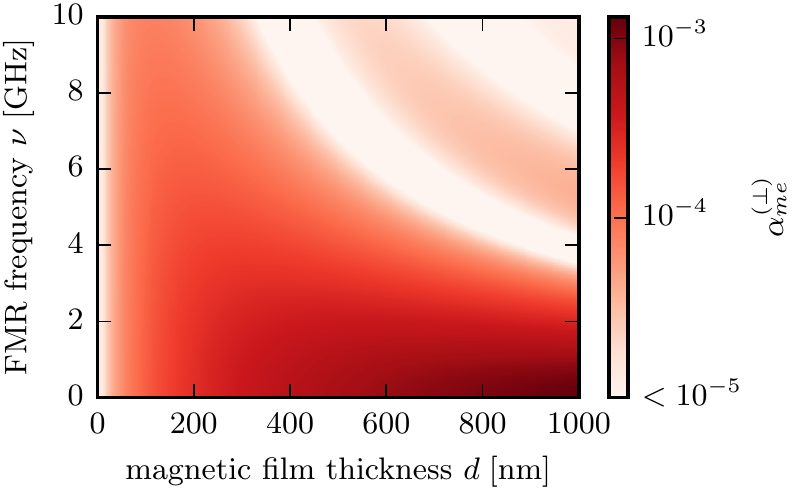}
\par\end{centering}
\caption{Phonon pumping-enhanced $\alpha_{me}^{(\perp)}$ in a YIG film sandwiched
between two infinitely thick GGG layers.\label{fig:alpha sandwich}}
\end{figure}

\begin{figure}
\begin{centering}
\includegraphics{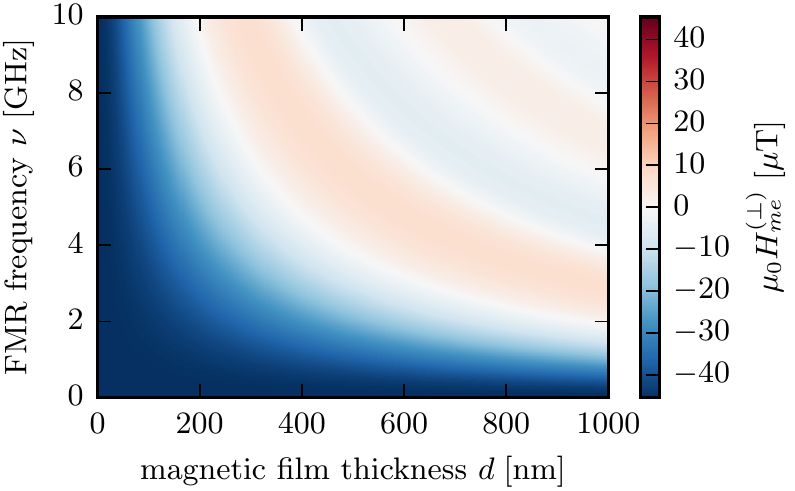}
\par\end{centering}
\caption{Phonon pumping effective field $H_{me}^{(\perp)}$ in a YIG film sandwiched
between two infinitely thick GGG layers.\label{fig:field sandwich}}
\end{figure}

When a non-magnetic material is attached at both sides of the magnet
and elastic waves leave the magnet at $z=0$ and $z=-d$, the boundary
condition are
\begin{align}
j_{\pm}(-d^{-})-j_{\pm}(-d^{+}) & =\left(B_{\perp}-K_{1}\right)m_{\pm},\\
j_{\pm}(0^{+})-j_{\pm}(0^{-}) & =-\left(B_{\perp}-K_{1}\right)m_{\pm},\\
u_{\pm}(0^{+}) & =u_{\pm}(0^{-}),\\
u_{\pm}(-d^{+}) & =u_{\pm}(-d^{-}),
\end{align}
with $d^{\pm}=d\pm0^{+}$. Since the Hamiltonian is piece-wise constant
\begin{equation}
u_{\pm}(z,t)=\begin{cases}
C_{\pm}e^{ikz-i\omega t}, & z>0\\
D_{\pm}e^{i\tilde{k}z-i\omega t}+E_{\pm}e^{-i\tilde{k}z-i\omega t}, & -d<z<0\\
F_{\pm}e^{-ikz-i\omega t}. & z<-d
\end{cases},
\end{equation}
Using the boundary conditions
\begin{equation}
v=\frac{u_{\pm}(0)-u_{\pm}(-d)}{dm_{\pm}}=\frac{M_{s}\omega_{c}}{\omega\gamma d\tilde{\rho}\tilde{c}_{t}}\frac{2}{i\frac{\rho c_{t}}{\tilde{\rho}\tilde{c}_{t}}-\cot(\frac{\tilde{k}d}{2})},
\end{equation}
leading to the damping coefficient
\begin{equation}
\alpha_{me}^{(\perp)}=\mathrm{\left(\frac{\omega_{c}}{\omega}\right)^{2}}\frac{M_{s}}{\gamma d\tilde{\rho}\tilde{c}_{t}}\frac{2}{\frac{\rho c_{t}}{\tilde{\rho}\tilde{c}_{t}}+\frac{\tilde{\rho}\tilde{c}_{t}}{\rho c_{t}}\cot^{2}\left(\frac{\tilde{k}d}{2}\right)}.
\end{equation}
When $\tilde{\rho}\tilde{c}_{t}=\rho c_{t}$,
\begin{equation}
\alpha_{me}^{(\perp)}=\mathrm{\left(\frac{\omega_{c}}{\omega}\right)^{2}}\frac{2M_{s}}{\gamma d\rho c_{t}}\sin^{2}\left(\frac{\tilde{k}d}{2}\right),
\end{equation}
which differs from the $\sin^{4}\left(\tilde{k}d/2\right)$ dependence
obtained for the magnet|non-magnet bilayer. This result can be explained
by the phonon angular momentum leaking at two interfaces that should
enhance the damping for thin magnetic films. However, the phonon pumping
is a coherent process that couples both interfaces, so the damping
is not increased simply by a factor of 2 as in case of incoherent
spin pumping of a magnetic film sandwiched by metals. The position
of the resonances, $\tilde{k}d=(2n+1)\pi/2$ with $n\in\mathbb{N}_{0}$,
are independent of the ratio $\rho c_{t}/\tilde{\rho}\tilde{c}_{t}$
with 
\begin{equation}
\alpha_{me}=\mathrm{\left(\frac{\omega_{c}}{\omega}\right)^{2}}\frac{2M_{s}}{\gamma d\rho c_{t}}.
\end{equation}
$\alpha_{me}^{(\perp)}$ and the effective magnetic field for YIG
sandwiched between two infinitely thick GGG layers are shown in Figs.~\ref{fig:alpha sandwich}
and \ref{fig:field sandwich}. 

\section{Flexural (bending) waves in thin beams\label{sec:Thin-beams}}

\begin{figure}
\begin{centering}
\includegraphics[scale=1.5]{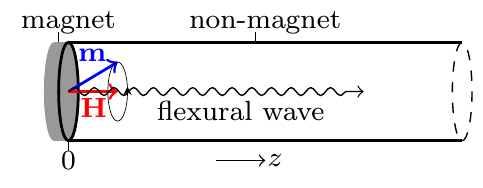}
\par\end{centering}
\caption{Thin film magnet (shaded) with magnetization $\mathbf{m}$ attached
to a thin semi-infinite elastic beam.\label{fig:beam}}
\end{figure}

In the main text we focus on the generation of transverse or longitudinal
sound waves. In free-standing structured samples such as cantilevers,
additional modes become important that can be excited by magnetization
dynamics as well. This can be illustrated by a thin cylindrical elastic
beam (see Fig.~\ref{fig:beam}) with cross section area $A=\pi r^{2}$,
in which flexural waves are generated by the magnet of volume $V=Ad$
attached to the top of the beam. The elastic energy according to the
Euler-Bernoulli beam theory \cite{Landau1970}
\begin{equation}
E_{el}=\int_{0}^{L}dz\left[\frac{1}{2}\rho A\dot{\mathbf{u}}^{2}(z,t)+\frac{1}{2}E_{Y}I_{\perp}\mathbf{u}''^{2}(z,t)\right],
\end{equation}
leads to the equation of motion for the flexural waves \cite{Landau1970}
\begin{equation}
\rho A\ddot{u}_{\pm}(z,t)+E_{Y}I_{\perp}u_{\pm}^{(4)}(z,t)=0,
\end{equation}
where $I_{\perp}=\int dA\,x^{2}=\pi r^{4}/4$ and elastic modulus
$E_{Y}=\mu(3\lambda+2\mu)/\left(\lambda+\mu\right)$. The dispersion
relation of the flexural waves is quadratic,
\begin{equation}
\omega=\sqrt{\frac{E_{Y}I_{\perp}}{\rho A}}k^{2}.
\end{equation}
When the dimensions of the magnet are much smaller than the wavelength
of the elastic waves, the magnetoelastic coupling is suppressed and
magnetization and lattice are coupled exclusively by the magnetocrystalline
and, in contrast to the bulk magnet, also the shape anisotropies,
\begin{equation}
E_{A}=E_{K}+E_{D},
\end{equation}
with
\begin{align}
E_{K} & =K_{1}V(\mathbf{m}\times\mathbf{n})^{2},\\
E_{D} & =\frac{1}{2}\mu_{0}V\mathbf{M^{\mathit{T}}}\mathcal{D}\mathbf{M}.
\end{align}
For a thin magnetic film $\mathcal{D}_{zz}=D_{3}=1$ . When the magnet
become very thick ($d\gg r$), i.e. a needle with its point forming
the contact, and a coordinate system with $z$ axis along the surface
normal $\mathbf{n}$, $\mathcal{D}_{xx}=\mathcal{D}_{yy}=D_{\perp}=-1/2$.
All other $\mathcal{D}_{\alpha\beta}$ vanish. In contrast to the
extended bilayer treated in the main text $\mathbf{n}$ is now a dynamic
variable with $n_{\pm}=-u_{\pm}'(0,t)$. The mechanical torque exerted
by the magnet on the elastic beam reads
\begin{equation}
\boldsymbol{\tau}=\dot{\mathbf{L}}=\frac{VM_{s}}{\gamma}\dot{\mathbf{m}}+\dot{{\bf J}},\label{eq:mechanical torque}
\end{equation}
where $\dot{{\bf J}}=\mu_{0}VM_{s}\mathbf{m}\times{\bf H}_{\mathrm{ext}}$
is the torque exerted by the external magnetic field on the total
angular momentum. For a magnet with equilibrium magnetization $\mathbf{m}\parallel\mathbf{n}_{0}$

\begin{equation}
\tau_{\pm}=\pm if\left(m_{\pm}-n_{\pm}\right),\label{eq:torque}
\end{equation}
where $f=VM_{s}\omega_{A}/\gamma$ and 
\begin{equation}
\omega_{A}=\begin{cases}
2\gamma K_{1}/M_{s}-\gamma\mu_{0}M_{s}, & \textrm{thin film}\\
2\gamma K_{1}/M_{s}+\frac{1}{2}\gamma\mu_{0}M_{s}, & \textrm{needle}
\end{cases}.
\end{equation}
In order to compute the angular momentum current pumped into the attached
non-magnet, $\dot{L}_{\pm}=\tau_{\pm}$, we have to specify four boundary
conditions. Two are provided by the assumption that the beam is infinitely
long so that the are no reflections. The absence of shear forces at
the boundary is expressed by $u'''(0,t)=0$, \textcolor{black}{while
the bending by the torque follows from the principle of least action
\cite{Magrab2012} }

\begin{align}
u_{\pm}''(0,t)= & \pm i\frac{\tau_{\pm}}{E_{Y}I_{\perp}}.\label{eq:boundary condition}
\end{align}
The general solution for the differential equation can be written
\begin{equation}
u_{\pm}(z,t)=e^{-i\omega t}\left(A_{\pm}e^{ikz}+B_{\pm}e^{-kz}\right),
\end{equation}
because there are no back-reflections in the semi-infinite beam. We
find
\[
n_{\pm}=wm_{\pm}
\]
with
\begin{equation}
w=\frac{-2f}{E_{Y}I_{\perp}k}\left(1+i-\frac{2f}{E_{Y}I_{\perp}k}\right)^{-1},
\end{equation}
and the following source term in the LLG equation,
\begin{align}
\left.\dot{m}_{\pm}\right|_{an} & =\mp i\omega_{A}n_{\pm}\nonumber \\
 & =\pm i\omega_{A}\mathrm{Re}(w)m_{\pm}\mp\omega_{A}\mathrm{Im}(w)m_{\pm},
\end{align}
The first term on the right-hand-side is a field-like torque equivalent
to the effective field
\begin{equation}
\mu_{0}\mathbf{H}_{an}=\frac{\omega_{A}}{\gamma}\mathrm{Re}(w)\mathbf{e}_{z},\label{eq:H_an}
\end{equation}
and the second one a damping-like torque with damping coefficient
\begin{equation}
\alpha_{an}=-\frac{\omega_{A}}{\omega}\mathrm{Im}\,w.\label{eq:alpha_an}
\end{equation}
Since for weak magnetoelastic coupling we expect $\alpha_{an}\ll1$
and therefore $|w|\ll1$, which is equivalent to $2f/(E_{Y}I_{\perp}k)\ll1$,
we may approximate 
\begin{align}
w & \approx\frac{f\left(1-i\right)}{E_{Y}I_{\perp}k},\\
\alpha_{an} & \approx\frac{VM_{s}\omega_{A}^{2}}{\omega k\gamma I_{\perp}E_{Y}},\\
\mu_{0}\mathbf{H}_{an} & \approx-\frac{VM_{s}\omega_{A}^{2}}{\gamma^{2}E_{Y}I_{\perp}k}\mathbf{e}_{z}.
\end{align}
The damping enhancement scales as
\begin{equation}
\alpha_{an}\propto\frac{V}{A^{2}\omega^{\frac{3}{2}}}.
\end{equation}
For a needle-shaped YIG magnet attached to GGG with $E_{Y}=2.5\times10^{11}\;\mathrm{Pa}$
and $\omega_{A}=1.4\times10^{10}\;\mathrm{s^{-1}}$ 
\begin{align}
\alpha_{an} & \approx8.6\times10^{-6}\;\frac{d/\mathrm{nm}}{(\nu/\mathrm{GHz})^{\frac{3}{2}}(r/\mathrm{nm})^{2}},\\
\mu_{0}\left|\mathbf{H}_{an}\right| & \approx3.1\times10^{-7}\;\frac{d/\mathrm{nm}}{(\nu/\mathrm{GHz})^{\frac{1}{2}}(r/\mathrm{nm})^{2}}\;\mathrm{T},
\end{align}
which are very small numbers even at nanoscale dimensions. 
\end{document}